\documentclass[graybox]{svmult}

\usepackage{mathptmx}       % selects Times Roman as basic font
\usepackage{helvet}         % selects Helvetica as sans-serif font
\usepackage{courier}        % selects Courier as typewriter font
\usepackage{type1cm}        % activate if the above 3 fonts are
\usepackage{makeidx}         % allows index generation
\usepackage{graphicx}        % standard LaTeX graphics tool
\usepackage{multicol}        % used for the two-column index
\usepackage[bottom]{footmisc}% places footnotes at page bottom

\makeindex             % used for the subject index
\begin{document}

\title*{The new Kepler picture of variability among A and F type stars}
\author{K. Uytterhoeven and KASC WG\#10}
%\authorrunning{Uytterhoeven et al.}
\institute{K. Uytterhoeven  \at 1)  Instituto de Astrof\'{\i}sica de Canarias (IAC), Tenerife; 2) Dept. Astrof\'{\i}sica, Universidad de La Laguna (ULL), Tenerife; 3) Laboratoire AIM, CEA/DSM-CNRS-Universit\'e Paris Diderot; CEA, IRFU, SAp, Centre de Saclay; 4) Kiepenheuer-Institut f\"ur Sonnenphysik, Freiburg im Breisgau \email{katrien@iac.es}
}
\maketitle
% Please use both starred abstract and non-starred abstract.

\abstract*{The {\it Kepler} spacecraft is providing photometric time series with micro-magnitude precision for thousands of variable stars. The continuous  time-series of unprecedented time span  open up opportunities to study the pulsational variability in much more detail than was previously possible from the ground.  We present a first general characterization of the variability  of A-F type stars as observed in the  {\it Kepler} light curves of a sample of 750 candidate A-F type stars, and  investigate the relation between $\gamma$\,Doradus, $\delta$\,Scuti, and hybrid stars. Our results imply an investigation of pulsation mechanisms to supplement the $\kappa$ mechanism and convective blocking effect to drive hybrid pulsations and suggest a revision of the current observational instability strips of delta Scuti and gamma Doradus stars in case the currently available values of $T_{\rm eff}$ and $\log g$ will be confirmed.}

\abstract{The {\it Kepler} spacecraft is providing photometric time series with micro-magnitude precision for thousands of variable stars, what opens up opportunities to study the pulsational variability in much more detail than was previously possible from the ground.  We present a first general characterization of the variability  of A-F type stars as observed in the  {\it Kepler} light curves of a sample of 750 candidate A-F type stars.}

\section{Motivation}
\label{sec:1}
The classes of $\delta$\,Scuti ($\delta$\,Sct) and $\gamma$\,Doradus ($\gamma$\,Dor) stars lie next to each other along the main sequence (MS) in the Hertzsprung-Russell diagram (HR-diagram), with a small overlap of their instability strips (IS). The two classes have different pulsational properties (see the description below). It remains unclear if there exists a relation or link between them. Key objects for such an investigation are hybrid $\delta$\,Sct / $\gamma$\,Dor stars. Based on solely ground-based observations only a couple of hybrid stars have been discovered (e.g. \cite{HF05}, \cite{KU2008}). Thanks to the long-term monitoring with  photometric precision at the level of micro-magnitudes of space missions such as MOST, CoRoT, and {\it Kepler}, many more hybrid A-F type pulsators are being discovered (\cite{Grig2010}). In particular, the {\it Kepler} mission is providing high-quality time series for several hundreds of A-F type stars, allowing for the first time a comprehensive analysis of the variability of a large sample of A-F type stars.

\begin{svgraybox}
{\bf $\delta$\,Sct stars:} Stars that pulsate in low-order gravity (g) and pressure (p) modes with periods between 15\,min and 5\,h (frequencies $>$ 5 d$^{-1}$). Modes are excited through the  $\kappa$-mechanism. Typical global parameters are:  $\log g=3.2-4.3$ and $T_{\rm eff}= 6300-8600$\,K (\cite{catalogR00}).
\\
{\bf $\gamma$\,Dor stars:} Stars that pulsate in high-order g modes with typical periods between 8~h and 3~d (frequencies $<$ 5 d$^{-1}$). Modes are excited by a flux modulation mechanism induced by the upper convective layer (\cite{Guzik2000}). Typical global parameters are:  $\log g=3.9-4.3$ and $T_{\rm eff}= 6900-7500$\,K (\cite{Handler2002}).
\end{svgraybox}

We describe here the main results of an analysis of {\it Kepler} survey phase data (May 2009 -- March 2010) of a sample of 750 candidate A-F type stars, as presented by Uytterhoeven et al. (\cite{KU2011}). The sample consists of stars initially assigned to KASC\footnote{Kepler Asteroseismic Science Consortium, http://astro.phys.au.dk/KASC} Working Groups on $\gamma$\,Dor and $\delta$\,Sct stars, to which  stars that clearly show $\delta$\,Sct or $\gamma$\,Dor type pulsations were added. 

\section{Observational classification}
We investigated the light curves, the extracted frequency spectra, and the detected frequencies of the 750 sample stars, with the aim to classify candidate $\gamma$\,Dor, $\delta$\,Sct, and hybrid stars. We used an automatic procedure to identify obvious harmonic and alias frequencies, which were subsequently discarded. We stress that only seemingly 'independent' frequencies with amplitudes above 20ppm were considered in the analysis described below. We compiled a database of available values of physical parameters such as $T_{\rm eff}$ and $\log g$. For most stars the only available source was the Kepler Input Catalogue (\cite{KIC}). For about 110 stars independent  $T_{\rm eff}$ and $\log g$ values were available in the literature or were derived from new ground-based data. 

We classified the A-F type stars in three groups: $\delta$\,Sct stars,  $\gamma$\,Dor stars, and hybrid stars. The classification criteria used were: a $\delta$\,Sct star shows mainly frequencies $> 5$d$^{-1}$; a $\gamma$\,Dor star shows mainly frequencies $< 5$d$^{-1}$; a star was considered a candidate hybrid star when at least two independent frequencies with amplitudes above 100ppm were detected in both domains, whereby the highest amplitudes in the two domains are comparable or do not differ more than a factor of 5--7. We tried to avoid misclassifications due to long-term signals caused by rotation or short-term signals introduced by the spectral window. Using these criteria, we classified 27\% of the sample stars as candidate $\delta$\,Sct star, 13\% as candidate $\gamma$\,Dor star, and 23\% (171 stars) as candidate hybrid star. {\it These numbers imply that hybrid $\delta$\,Sct / $\gamma$\,Dor pulsators are not uncommon.}

The remaining 275 stars in our sample were identified as rotationally modulated/active stars, binaries, stars of different spectral type, or stars that show no clear periodic variability.

\section{Observational results}
We investigated physical properties, such as $T_{\rm eff}$ and $\log g$, of the 
 475 {\it Kepler} stars classified as $\delta$\,Sct, $\gamma$\,Dor, or hybrid star, and characterized them in terms of number of independent frequencies, amplitude heights, and range of detected frequencies.

\begin{figure}
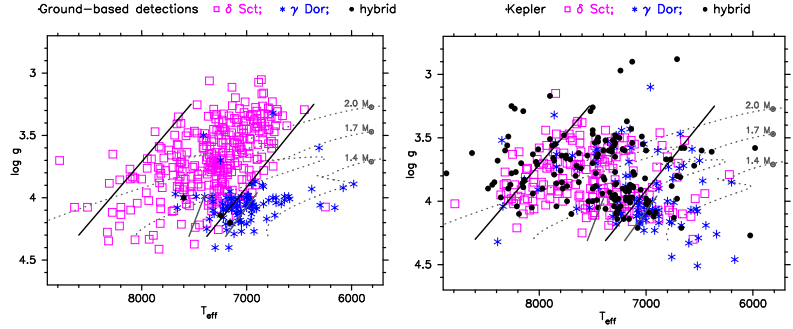

\centering
\begin{tabular}{cc}
\resizebox{0.44\linewidth}{!}{\rotatebox{-90}{\includegraphics{Uytterhoeven1_fig1.eps}}} &
\resizebox{0.44\linewidth}{!}{\rotatebox{-90}{\includegraphics{Uytterhoeven1_fig2.eps}}} \\
\end{tabular}
\caption{(left): ($T_{\rm eff}$, $\log g$)-diagram of $\delta$\,Sct, $\gamma$\,Dor, and hybrid stars detected from the ground. (right): ($T_{\rm eff}$, $\log g$)-diagram of the 475 {\it Kepler} stars classified as A-F type pulsators. The solid thick black and light grey  lines mark the blue and red edge of the observed IS of $\delta$\,Sct\,and $\gamma$\,Dor\,stars (\cite{catalogR00},\cite{Handler2002}). The dotted lines mark evolutionary tracks for MS stars with different masses.}
\label{logTefflogg}
\end{figure}

The right panel of Fig.\,\ref{logTefflogg} shows the position of the 475 {\it Kepler} stars in the  ($T_{\rm eff}$, $\log g$)-diagram, together with the observational IS for $\delta$\,Sct and $\gamma$\,Dor stars (\cite{catalogR00},\cite{Handler2002}). Squares, asterisks and bullets represent $\delta$\,Sct, $\gamma$\,Dor, and hybrid stars, respectively. As a comparison, the left panel of Fig.\,\ref{logTefflogg} shows the  $\delta$\,Sct, $\gamma$\,Dor, and hybrid stars known from before the space mission and detected from ground-based observations. 

Based on the currently available values of $T_{\rm eff}$ and $\log g$, and taking into account their (large) error bars, we find that:
\begin{itemize}
\item $\gamma$\,Dor pulsations are detected outside the $\gamma$\,Dor IS, in both cooler and hotter stars than previously observed from the ground;
\item $\delta$\,Sct pulsators are detected beyond the red edge of the $\delta$\,Sct IS;
\item hybrid stars are not confined to the small overlapping region of the $\delta$\,Sct and $\gamma$\,Dor IS, and are detected in the entire region between the blue edge of the $\delta$\,Sct IS and the red edge of the $\gamma$\,Dor IS, and beyond.
\end{itemize}
{\it These results seem to suggest that the edges of the so far accepted observational IS need to be revised.} However, we need accurate values of $T_{\rm eff}$ and $\log g$ for all stars to confirm this finding.  
Also, the known driving mechanisms do not explain the observed hybrid pulsations in hot and cool A-F type stars. {\it Alternative pulsation mechanisms need to be investigated to supplement the $\kappa$\,mechanism and convective blocking effect to drive hybrid pulsations.}

Another finding is that there is a great concentration of stars near the overlap of the $\delta$\,Sct and $\gamma$\,Dor IS: {\it hybrid, $\delta$\,Sct, and $\gamma$\,Dor stars coincide in the same region of the HR-diagram.} The fact that stars with similar temperatures show a different pulsational behaviour raises questions, e.g., What makes a star a $\delta$\,Sct, $\gamma$\,Dor, or hybrid pulsator? What is the difference in their internal physics? 

When investigating the number of independent frequencies, the associated amplitudes, and the observed frequency range for the three groups of A-F type pulsators, we find:
\begin{itemize}
\item Up to 500 non-combination frequencies are detected in the variability of the {\it Kepler} light curves of A-F type pulsators. However, the majority (66\%) shows less than 100 independent variable signals;
\item The number of detected frequencies versus $T_{\rm eff}$\,follows a distribution that peaks between the centre and the red edge of the respective IS;
\item For about 59\% of the stars the highest amplitude is lower than 2000\,ppm. In general, higher amplitudes are detected in $\delta$\,Sct stars than in $\gamma$\,Dor stars.
\item The 'frequency gap' between 5--10 d$^{-1}$ predicted by the current instability models for hybrid stars is not observed in the {\it Kepler} data.
\end{itemize}

To summarize, {\it hybrid  $\delta$\,Sct /$\gamma$\,Dor pulsations seem to be more common than previously anticipated} and are found in stars with a large range of temperatures. These results open up new questions and demand a further investigation of the pulsational driving mechanisms in A-F type stars. Thanks to the high-quality {\it Kepler} data of hundreds of A-F type stars that are becoming available we have good prospects in making progress in this field.

\begin{acknowledgement}
KU acknowledges financial support by the Deutsche Forschungsgemeinschaft (DFG) in the framework of project UY 52/1-1, and by the Spanish National Plan of R\&D for 2010, project AYA2010-17803. 
\end{acknowledgement}

\end{document}